\documentclass[aps,pra,reprint,groupedaddress]{revtex4-2}

\usepackage{graphicx}
\usepackage{amsfonts}
\usepackage{amsmath}
\usepackage{amssymb}
\usepackage{braket}
\usepackage{mathtools}
\usepackage{siunitx}
\usepackage{accents}
\usepackage[utf8]{inputenc}
\usepackage{circledsteps}
\usepackage{xcolor}
\usepackage{enumerate}
\usepackage{mathtools}
\usepackage{cases}
\usepackage{bm}
\usepackage{empheq}
\usepackage[makeroom]{cancel}
\usepackage{comment}
\usepackage{hyperref}

\begin{document}

\title[Quantum Synchronization Circuits]{Quantum Synchronization in Nonconservative Electrical Circuits with Kirchhoff-Heisenberg Equations}

\author{M.~Mariantoni}
\email[Corresponding author: ]{matteo.mariantoni@uwaterloo.ca}
\affiliation{Institute for Quantum Computing, University of Waterloo, 200 University Avenue West, Waterloo, Ontario N2L 3G1, Canada}
\affiliation{Department of Physics and Astronomy, University of Waterloo, 200 University Avenue West, Waterloo, Ontario N2L 3G1, Canada}

\author{N.~Gorgichuk}
\affiliation{Institute for Quantum Computing, University of Waterloo, 200 University Avenue West, Waterloo, Ontario N2L 3G1, Canada}
\affiliation{Department of Physics and Astronomy, University of Waterloo, 200 University Avenue West, Waterloo, Ontario N2L 3G1, Canada}

\date{\today}

\begin{abstract}
We investigate quantum synchronization phenomena in electrical circuits that incorporate specifically designed nonconservative elements. A dissipative theory of classical and quantized electrical circuits is developed based on the Rayleigh dissipation function. The introduction of this framework enables the formulation of a generalized version of classical Poisson brackets, which are termed Poisson-Rayleigh brackets. By using these brackets, we are able to derive the equations of motion for a given circuit. Remarkably, these equations are found to correspond to Kirchhoff's current laws when Kirchhoff's voltage laws are employed to impose topological constraints, and vice versa. In the quantum setting, the equations of motion are referred to as the Kirchhoff-Heisenberg equations, as they represent Kirchhoff's laws within the Heisenberg picture. These Kirchhoff-Heisenberg equations, serving as the native equations for an electrical circuit, can be used in place of the more abstract master equations in Lindblad form. Furthermore, without employing a slowly varying envelope approximation, we directly solve the Kirchhoff-Heisenberg equations for the complete quantized canonical coordinates~$\hat{\phi}$ and $\hat{q}$ in presence of arbitrary dissipation, rather than the split bosonic operators~$\hat{a}$ or $\hat{a}^{\dag}$ for weak dissipation. To validate our theoretical framework, we examine three distinct circuits. The first circuit consists of two resonators coupled via a nonconservative element. The second circuit extends the first to incorporate weakly nonlinear resonators, such as transmons. Lastly, we investigate a circuit involving two resonators connected through an inductor in series with a resistor. This last circuit, which incidentally represents a realistic implementation, allows for the study of a singular system, where the absence of a coordinate leads to an ill-defined system of Hamilton's equations. To analyze such a pathological circuit, we introduce the concept of auxiliary circuit element. After resolving the singularity, we demonstrate that this element can be effectively eliminated at the conclusion of the analysis, recuperating the original circuit. We show that quantum synchronization persists even in the case of a pathological circuit.
\end{abstract}


\maketitle

\section{INTRODUCTION \label{Sec::INTRODUCTION}}

Understanding and modeling nonconservative forces, such as friction, presents a significant challenge in the field of physics. In classical mechanics, basic models of friction can be easily integrated into the Newtonian framework. However, more sophisticated theories, such as the Lagrangian and Hamiltonian formalism, necessitate the introduction of the Rayleigh dissipation function to account for generalized forces associated with velocity-dependent isotropic friction~\cite{Cercignani:1976}. While some researchers regard the dissipation function as an~\textit{ad hoc} tool within Hamiltonian mechanics, it serves as a mathematically elegant extension of Newtonian friction to align with Hamilton's formalism. In essence, the dissipation function adds a layer of abstraction in the case of nonconservative forces akin to the Hamilton's function for conservative forces.

Mechanical systems exhibit striking parallels with general nonlinear and nonconservative electrical circuits. In this analogy, Newton's laws in mechanics assume the role of Kirchhoff's laws for circuits. These laws, in their most comprehensive form, are derived from the time-dependent Maxwell's equations, particularly Faraday's law of induction and Amp\`{e}re-Maxwell's equation, which give rise to Kirchhoff's voltage law~(KVL) and Kirchhoff's current law~(KCL). The dual nature of Kirchhoff's laws renders them effective for both enforcing a circuit's constraints and deriving its equations of motion~(EOMs). Typically, KVLs set the constraints for the circuit's fluxes (and voltages), with a combination of KCLs and the constitutive relationships of the circuit elements yielding the~EOMs for the charges (and currents), or vice versa. Fluxes~$\phi$ and charges~$q$ are assumed to be the circuit's canonical coordinates.

A pivotal step towards establishing a refined formalism for analyzing arbitrary nonlinear and nonconservative~$RLC$ classical circuits was made in~$1964$ by Brayton and Moser~\cite{Brayton:1964}. Their groundwork eventually culminated in a Hamiltonian formulation for complete nonlinear~$RLC$ networks, as demonstrated by~Weiss and Mathis in the work of Ref.~\cite{Weiss:1997}. This formulation explicitly uncovers the linkage between the Brayton-Moser equations and Hamiltonian formalism for dissipative systems.

In this paper, we extend the dissipative theory proposed by Weiss and Mathis by introducing the concept of \emph{Poisson-Rayleigh brackets}. These brackets, which are a generalization of the traditional Poisson brackets to include derivatives of canonical coordinates, enable us to express the~EOMs of a nonconservative circuit in a concise manner similar to Hamilton's equations. Notably, the nonconservative~EOMs derived using the Poisson-Rayleigh brackets represent a first-order system of equations that is equivalent to the second-order system obtained directly from Kirchhoff's laws.

Subsequently, we proceed to explore the quantized version of our theory by replacing the Poisson-Rayleigh brackets with commutators, following the standard canonical quantization procedure. The quantized EOMs, with the canonical coordinates~$\phi\rightarrow\hat{\phi}$ and $q\rightarrow\hat{q}$, are found to be identical to their classical counterparts. We refer to this set of equations as the \emph{Kirchhoff-Heisenberg equations}, as they represent the quantized form of Kirchhoff's laws for the time-dependent observables~$\hat{\phi}(t)$ and $\hat{q}(t)$, corresponding to the quantized canonical coordinates in the Heisenberg picture. Since we consider only linear and weakly nonlinear circuits, the canonical coordinates are quantized by means of bosonic creation and annihilation operators of harmonic oscillators, $\hat{a}^{\dag}$ and $\hat{a}$~\cite{Yurke:2004}.

Our analysis is exclusively conducted in the Heisenberg picture, which is deemed the most appropriate framework for describing the observables of the circuit in accordance with Kirchhoff's laws. Notably, we refrain from employing the commonly used slowly varying envelope approximation introduced by Yurke in the work of Ref.~\cite{Yurke:2004}. This approximation simplifies a system of second-order quantized Kirchhoff's laws involving~$\hat{\phi}$ and $\hat{q}$ to a first-order system involving either~$\hat{a}^{\dag}$ or $\hat{a}$, but is valid only under weak dissipation conditions. Instead, we solve the quantized~EOMs for the complete normalized observables, encompassing both rotating and counter-rotating terms, $\hat{\phi}=\hat{a}^{\dag}+\hat{a}$ and $\hat{q}=\left(\hat{a}-\hat{a}^{\dag}\right)/i$. This approach, which is not merely a way to avoid a rotating-wave approximation, allows us to address problems with arbitrary degrees of dissipation. This feature may be uninteresting when studying systems that are designed to have low loss (e.g., superconducting qubits), but it proves invaluable when nonconservative elements are deliberately incorporated to play an active role in the circuit design.

A test case involving the use of nonconservative elements as a resource rather than an impediment in quantum dynamics is explored through the passive synchronization of coupled circuits. In its simplest form, circuit synchronization focuses on two identical lossless~$LC$ resonators connected via a resistor~$R$. When suitably initialized, the charge and flux observables of the two resonators, $q_{1,2}$ or $\phi_{1,2}$, exhibit complete amplitude and phase synchronization after a transient period determined by the time constant~$RC$.

Expanding upon Barbi's classical circuit design presented in the work of Ref.~\cite{Barbi:2021}, our research incorporates local loss in each resonator and introduces inductive and capacitive coupling alongside the original resistive coupling. The Kirchhoff-Heisenberg equations are derived to analyze the circuits and explore various regimes in both classical and quantum settings, including the resistive coupling of two nonlinear circuits such as a pair of transmons~\cite{Koch:2007}. Our methods represent a significant extension of Barbi's classical resonator theory. Results demonstrate the persistence of quantum synchronization even in the presence of high local loss and detuned resonators.

Our study of quantum synchronization in circuit quantum electrodynamics~(QED) benefits from comparing it to similar studies in the field. The dissipative interaction between two qubits has been researched by Cattaneo and Militello, as described in their respective works of Ref.~\cite{Cattaneo:2021} and \cite{Militello:2021}.

In Cattaneo's work, the authors analyze the coupling between two transmons that are capacitively coupled to a common resistor, while Militello investigates a pair of qubits coupled to a dissipating resonator. Both studies explore the nonconservative elements of the circuits and their impact on synchronization dynamics through the solution of a master equation in Lindblad form in the Schr\"{o}dinger picture.

However, while Lindbladians are commonly used in circuit~QED, they offer an indirect method that may not fully capture the topological and electrical characteristics of the original circuit. In contrast, our work introduces the Kirchhoff-Heisenberg equations as a native circuital approach. These equations can be derived directly from the electrical circuit diagram, avoiding the need for intermediate steps or approximations. This approach provides a more accurate representation of the circuit's behavior and enhances our understanding of quantum circuit theory in general, and quantum synchronization dynamics in particular. We find it advantageous to work within the Heisenberg picture using Kirchhoff-Heisenberg equations, which closely align with circuit theory, as opposed to employing Lindbladians, at the cost of not being able to directly access the system's density matrix.

In this study, we propose a circuit design to bridge the gap between theoretical concepts and experimental implementation. Specifically, we suggest coupling two resonators (or two transmons) using a \emph{realistic resistor} in the form of a long conducting strip, which can be modeled as a resistor and a parasitic inductor in series (referred to as an~$RL$-series coupler). Interestingly, this unique coupling arrangement results in a \emph{pathological circuit}, characterized by an incomplete set of canonical coordinates within its Hamiltonian due to the presence of a nonconservative element--the resistor. The circuit is said to be singular as it lacks the Hamilton's equation corresponding to the missing coordinate. This singularity, which is attributable to the resistor modifying the circuit's topology, would not be present if a purely inductive or capacitive coupling were used instead.

Singular electrical circuits, such as the one proposed in this study, have garnered significant attention in recent research due to their pivotal role in furthering the development of a complete quantized theory. Previous works by Rymarz and DiVincenzo~\cite{Rymarz:2023} and Osborne~\textit{et al.}~\cite{Osborne:2023} have explored the challenges associated with singular Hamiltonians in superconducting conservative circuits and proposed methods to address and quantify the singularity. In the first study, the authors show the inadequacy of the well-known Dirac-Bergmann algorithm for singular Hamiltonians in the case of nearly singular superconducting circuits. In the second study, the singularity is dealt with accounting for parasitic circuit elements.

Indeed, it is important to highlight that every circuit element that leads to a singularity, such as the~$RL$-series coupler in our example, is paired with a parasitic element that can be used to remedy the singularity. In our case, the~$RL$-series coupler is accompanied by a parasitic capacitor that makes it possible to complete the circuit Hamiltonian, thus resolving the singularity. In our study, we generalize the notion of parasitic component by introducing the concept of an \emph{auxiliary circuit element} as a theoretical tool to effectively eliminate singularities in pathological circuits. By strategically placing this element in the circuit to remove the singularity and then subsequently considering limiting cases where the auxiliary element becomes either an open or a short circuit to effectively recuperate the original circuit, we can treat a variety of pathological circuits. The relevance of parasitic elements in circuits was mentioned in the early work by Vool and Devoret~\cite{Vool:2017}, explored by Mariantoni in the work of Ref.~\cite{Mariantoni:2021}, and refined in Osborne's theory~\cite{Osborne:2023}. In our study, we provide a practical example showcasing the existence of quantum synchronization in pathological circuits characterized by singularities.

The results of our Kirchhoff-Heisenberg equations theory are intriguing, as they raise questions about the necessity of Hamiltonians in studying the time evolution of quantum observables (flux and charge) in circuits. Our findings suggest that, for both regular and pathological linear and nonlinear nonconservative circuits, Hamiltonians may not be essential. Specifically, in the case of linear and weakly nonlinear systems, our theory indicates that the Kirchhoff-Heisenberg equations can be derived and quantized with one set of bosonic operators per degree of freedom, independently of Hamiltonians (or Lagrangians). It is clear that dismissing the importance of the Hamilton and Rayleigh functions in relation to electrical circuits would be a mistake. Rather, these functions serve as essential components in a sophisticated theoretical framework that builds upon fundamental laws like Newton's and Kirchhoff's. The development of a refined theory using these functions represents the culmination of a journey that allows for the exploration and resolution of critical issues within circuit theory, including but not limited to circuit stability.

The paper is organized as follows. In Sec.~\ref{Sec:METHODS}, we introduce the theoretical methods used to derive analytically and solve numerically the~EOMs of three different circuits. In Sec.~\ref{Sec::RESULTS}, we show our results for the three circuits introduced in the Methods' section. In Sec.~\ref{Sec::DISCUSSION}, we further elaborate a few key aspects of our findings in both the Methods' and Results' sections. Finally, in Sec.~\ref{Sec::CONCLUSIONS}, we summarize our work and outline possible extensions.

\section{METHODS \label{Sec:METHODS}}

In Sec.~\ref{Subsec::Circuit:Theory:of:Two:Resonators:Interacting:Via:a:Nonconservative:Coupler}, we present the circuit theory of two lossy~$RLC$ resonators interacting via a nonconservative~$RLC$ coupler. We introduce the Kirchhoff-Heisenberg circuit equations. In Sec.~\ref{Subsec::Extension:to:Transmons}, we extend the analysis to the case of two lossless transmons coupled by a resistor. In Sec.~\ref{Subsec::Nonconservative:Pathological:Circuit}, we study a pathological circuit comprising two~$LC$ resonators connected by an~$RL$-series coupler. Finally, in Sec.~\ref{Subsec::Numerical:Simulations}, we introduce the methodology to perform numerical simulations.

\subsection{Circuit Theory of Two Resonators Interacting Via a Nonconservative Coupler \label{Subsec::Circuit:Theory:of:Two:Resonators:Interacting:Via:a:Nonconservative:Coupler}}

\begin{figure}[t!]
	\centering
	\includegraphics[width=1.0\columnwidth]{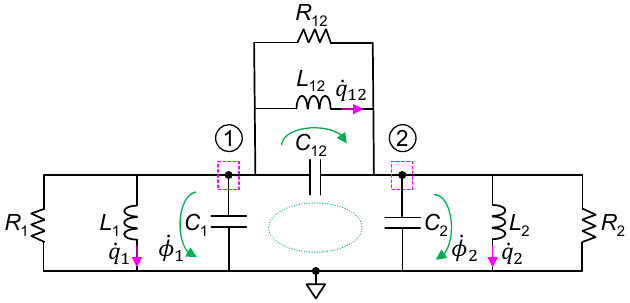}
	\caption{Circuit diagram of two lossy~$RLC$ resonators coupled by means of a nonconservative~$RLC$ circuit. We follow the standard convention that, on all passive circuit elements, voltages~[green (light gray) curved arrows] and currents~[magenta (middle gray) straight arrows] have the same direction. Additionally, we assume that a current entering a node is negative and a current exiting a node is positive. The dotted green~(light gray) ellipse in the center of the diagram indicates the capacitive loop~$C_1$-$C_{12}$-$C_2$, which allows us to find the voltage on~$C_{12}$ (see main text). The dashed magenta~(middle gray) rectangles indicate the independent nodes where~KCLs are applied in order to determine the circuit's~EOMs.}
		\label{Fig::figure01_res-coupl-circ-diagr_mt_mariantoni}
\end{figure}

Figure~\ref{Fig::figure01_res-coupl-circ-diagr_mt_mariantoni} displays two~\textit{RLC} resonators that are coupled through a nonconservative~\textit{RLC} circuit. The left~($k=1$) and right~($k=2$) resonators have resistance, inductance, and capacitance named~$C_k$, $L_k$, and $R_k$, respectively. In contrast, the coupler has these elements named~$C_{12}$, $L_{12}$, and $R_{12}$. The inclusion of the resistance~$R_k$ enables us to consider the loss due to each individual resonator, taking into account its \emph{local} properties. On the other hand, $R_{12}$ links the two resonators, representing a \emph{common} lossy environment. In the limiting scenario where the coupler is simplified to a resistor without reactive components (i.e., without any inductors and capacitors), the interaction between the left and right resonators occurs solely through a nonconservative element.

The circuit features~$N=3$ nodes and $B=9$ branches. Among these branches, $B^{(\text{c})}=7$ are associated with conservative elements (i.e., capacitors and inductors); the remaining~$B^{(\text{nc})}=2$ branches, on the other hand, are associated with nonconservative elements (i.e., resistors). Given that~$N-1=2<B-N+1=7$, it is convenient to apply~KVL to impose all the circuit's constraints. In fact, there are seven available loops in which to apply~KVL, resulting in two~EOMs. These~EOMs correspond to~KCL applied at the two independent nodes~$\Circled{1}$ and $\Circled{2}$ of the circuit~\footnote{For this simple circuit's topology, $N-1$ is equal to the circuit's degrees of freedom.}.

From~KVL (see Fig.~\ref{Fig::figure01_res-coupl-circ-diagr_mt_mariantoni}), it can be readily shown that all parallel branches have the same voltage. We select the voltages~$\accentset{\bullet}{\phi}_1$ and $\accentset{\bullet}{\phi}_2$ across~$C_1$ and $C_2$ as the circuit's independent variables. Applying~KVL around the capacitive loop~$C_1$-$C_{12}$-$C_2$, the voltage across~$C_{12}$ can be expressed as
\begin{equation}
	\accentset{\bullet}{\phi}_{12}=\accentset{\bullet}{\phi}_{1}-\accentset{\bullet}{\phi}_{2}.
		\label{Eq::KVL:7:phi12dot}
\end{equation}
This leads us to conclude that the flux across~$L_{12}$ can be denoted as
\begin{equation}
	\phi_{12}=\phi_{1}-\phi_{2}+\tilde{\phi},
		\label{Eq::KVL:7:phi12}
\end{equation}
where~$\tilde{\phi}$ is an arbitrary integration constant that can be conveniently set to zero, $\tilde{\phi}=0$. Additionally, we introduce the dependent variables~$\accentset{\bullet}{q}_1$, $\accentset{\bullet}{q}_{12}$, and $\accentset{\bullet}{q}_2$ to represent the currents flowing through~$L_1$, $L_{12}$, and $L_2$, respectively.

The total instantaneous power associated with the circuit can be expressed in terms of the voltages and currents across each branch, and can be written as follows:
\begin{eqnarray}
	\mathcal{P}(t)
	&=&\mathcal{P}^{(\textrm{c})}(t)+\mathcal{P}^{(\textrm{nc})}(t)
	\nonumber\\
	&=&\sum_{b=1}^{B^{(\text{c})}}\accentset{\bullet}{\phi}_{b}(t)\,
	\accentset{\bullet}{q}_{b}(t)+\sum_{b=1}^{B^{(\text{nc})}} \accentset{\bullet}{\phi}_{b}(t)\,\accentset{\bullet}{q}_{b}(t),
		\label{Eq::P:t}
\end{eqnarray}
where~$\mathcal{P}^{(\textrm{c})}$ represents the conservative power and $\mathcal{P}^{(\textrm{nc})}$ represents the nonconservative power. The constitutive relations to be used in Eq.~(\ref{Eq::P:t}) are~$\accentset{\bullet}{q}=C\accentset{\bullet\bullet}{\phi}$ for capacitors, $\accentset{\bullet}{\phi}=L\accentset{\bullet\bullet}{q}$ for inductors, and $\accentset{\bullet}{\phi}=R\accentset{\bullet}{q}$ for resistors.

The energy due to all conservative elements at any given time~$t$ can be obtained by integrating~$\mathcal{P}^{(\textrm{c})}(t)$ from the initial time~$t_0=0$ to $t$. Taking into account the constraint stated in Eq.~(\ref{Eq::KVL:7:phi12dot}), the expression for this energy is as follows:
\begin{eqnarray}
	\mathcal{E}(t)
	&=&\int_{0}^{t}dt'\,\mathcal{P}^{(\textrm{c})}(t')=\int_{0}^{t'}dt'\dfrac{d}{dt'}\Bigg\{
	\nonumber\\
	&{}&\dfrac{1}{2}C_1\accentset{\bullet}{\phi}_1^2(t')+\dfrac{1}{2}L_1 \accentset{\bullet}{q}_1^2(t')
	\nonumber\\
	&+&\dfrac{1}{2}C_{12}\left[ \accentset{\bullet}{\phi}_1(t')-\accentset{\bullet}{\phi}_2(t')\right]^2+ \dfrac{1}{2}L_{12}\accentset{\bullet}{q}_{12}^2(t')
	\nonumber\\
	&+&\dfrac{1}{2}C_2\accentset{\bullet}{\phi}_2^2(t')+\dfrac{1}{2}L_2 \accentset{\bullet}{q}_2^2(t')\Bigg\}.
		\label{Eq::E:t}
\end{eqnarray}
By setting all branch voltages and currents to zero at~$t_0$, the energy can be expressed as follows:
\begin{eqnarray}
	\mathcal{E}
	&=&\dfrac{\vec{\accentset{\bullet}{\phi}}^{\,\textrm{T}}
		\mathbf{C}
		\vec{\accentset{\bullet}{\phi}}}{2}
		+
		\dfrac{1}{2} L_1\accentset{\bullet}{q}_1^2+ \dfrac{1}{2}L_{12}\accentset{\bullet}{q}_{12}^2+\dfrac{1}{2} L_2\accentset{\bullet}{q}_2^2.
		\label{Eq::E}
\end{eqnarray}
In this equation, $\vec{\accentset{\bullet}{\phi}}^{\,\textrm{T}}=\left[\accentset{\bullet}{\phi}_1\,\,\accentset{\bullet}{\phi}_2\right]$,
\begin{equation}
	\mathbf{C}=
	\begin{bmatrix*}[c]
		\widetilde{C}_1 & -C_{12} \\
		-C_{12}         & \widetilde{C}_2
	\end{bmatrix*},
\end{equation}
and $\widetilde{C}_k=C_k+C_{12}$.

The circuit's energy~$\mathcal{E}$ is equivalent to the Hamiltonian~$\mathcal{H}$, but it is expressed in terms of two sets of generalized velocities~$\{\accentset{\bullet}{\phi}_1,\accentset{\bullet}{\phi}_2\}$ and $\{\accentset{\bullet}{q}_1,\accentset{\bullet}{q}_{12},\accentset{\bullet}{q}_{2}\}$ instead of one set of canonical coordinates~$\{\phi_1,\phi_2;q_1,q_2\}$. We can determine~$\mathcal{H}$ by solving the two systems of equations
\begin{subequations}
	\begin{empheq}[left=\empheqlbrace]{align}
		&\left\{
		\begin{array}{ll}
			\dfrac{\partial}{\partial\accentset{\bullet}{\phi}_1}\mathcal{E}
			&=\widetilde{C}_1\accentset{\bullet}{\phi}_1-C_{12} \accentset{\bullet}{\phi}_2=q_1
			\\[5.0mm]
			\dfrac{\partial}{\partial\accentset{\bullet}{\phi}_2}\mathcal{E}
			&=-C_{12} \accentset{\bullet}{\phi}_1+\widetilde{C}_2
			\accentset{\bullet}{\phi}_2=q_2
		\end{array}
		\right.
		\label{Subeq::EH:Transformation:q}
		\\[2.5mm]
		&\left\{
		\begin{array}{ll}
			\dfrac{\partial}{\partial\accentset{\bullet}{q}_1}\mathcal{E}
			&=L_1\accentset{\bullet}{q}_1=\phi_1
			\\[5.0mm]
			\dfrac{\partial}{\partial\accentset{\bullet}{q}_{12}} \mathcal{E}
			&=L_{12}\accentset{\bullet}{q}_{12}=\phi_1-\phi_2
			\\[5.0mm]
			\dfrac{\partial}{\partial\accentset{\bullet}{q}_2}\mathcal{E}
			&=L_2\accentset{\bullet}{q}_2=\phi_2
		\end{array}
		\right.
		\label{Subeq::EH:Transformation:phi}
	\end{empheq}
\end{subequations}
for the generalized velocities and substituting the results into Eq.~(\ref{Eq::E}). The constraint from Eq.~(\ref{Eq::KVL:7:phi12}) is accounted for in the second equation of the system of Eqs.~(\ref{Subeq::EH:Transformation:phi}).

It is worth noting that~$\mathcal{E}=\mathcal{T}_v(\accentset{\bullet}{\phi}_1,\accentset{\bullet}{\phi}_2)+\mathcal{T}_{\imath}(\accentset{\bullet}{q}_1,\accentset{\bullet}{q}_{12},\accentset{\bullet}{q}_2)$; that is, it is the sum of two kinetic energies: The kinetic energy~$\mathcal{T}_v$ associated with the circuit's voltages and the kinetic energy~$\mathcal{T}_{\imath}$ associated with its currents. This seemingly unorthodox expression for~$\mathcal{E}$ can be written in a more standard form as the sum of~$\mathcal{T}_v$ and a potential energy~$\mathcal{U}_{\phi}$, where~$\mathcal{U}_{\phi}$ is a function of~$\{\phi_1,\phi_2\}$; $\mathcal{U}_{\phi}$ can be obtained by solving the system of Eqs.~(\ref{Subeq::EH:Transformation:phi}) for the currents and substituting the results into the inductive terms of Eq.~(\ref{Eq::E})~\footnote{Typically, in most of the literature, $\mathcal{U}_{\phi}$ is given directly in terms of fluxes with the proper flux constraints; that is, the step associated with the inversion of Eqs.~(\ref{Subeq::EH:Transformation:phi}) is skipped.}. Since we are imposing~KVL constraints, the system of Eqs.~(\ref{Subeq::EH:Transformation:phi}) is associated with a diagonal matrix and, thus, it can be solved trivially (i.e., each equation in the system can be solved independently)~\footnote{If we were to use KCL constraints (in which case the EOMs would be KVL equations), then the capacitance matrix would be diagonal and trivially solvable.}.

In~$\mathcal{E}$, we are considering only the contribution from the conservative elements of the circuit; thus, $\partial\mathcal{E}/\partial\accentset{\bullet}{\phi}=\partial \mathcal{T}_v/\partial\accentset{\bullet}{\phi}=\partial\mathcal{L}/\partial\accentset{\bullet}{\phi}=q$, where~$\mathcal{L}$ is the system's Lagrangian (which we do not actually need to find!). The system of Eqs.~(\ref{Subeq::EH:Transformation:q}) therefore defines the conjugate momenta of the capacitor's voltages. This system cannot be solved trivially; it must be rewritten first in matrix form as
\begin{equation}
	\mathbf{C}\,\vec{\accentset{\bullet}{\phi}}=\vec{q},
		\label{Eq::C:vecphidot:vecq}
\end{equation}
where~$\vec{q}^{\,\textrm{T}}=\left[q_1\,\,q_2\right]$. By matrix inversion, we can readily obtain the generalized velocities~$\{\accentset{\bullet}{\phi}_1,\accentset{\bullet}{\phi}_2\}$ in terms of the canonical coordinates~$\{q_1,q_2\}$. More simply, $\vec{\accentset{\bullet}{\phi}}^{\,\textrm{T}}\mathbf{C}\vec{\accentset{\bullet}{\phi}}=\vec{q}^{\,\textrm{T}}\vec{q}/\mathbf{C}$. Since we impose the circuit's constraints using~KVL, each equation in the system of Eqs.~(\ref{Subeq::EH:Transformation:phi}) can be solved directly for~$\{\accentset{\bullet}{q}_1,\accentset{\bullet}{q}_{12},\accentset{\bullet}{q}_{2}\}$, allowing us to rewrite these three generalized velocities in terms of the two canonical coordinates~$\{{\phi}_1,{\phi}_{2}\}$. Following this procedure, the circuit Hamiltonian is finally obtained by rewriting~$\mathcal{E}$ in terms of the canonical coordinates:
\begin{equation}
	\mathcal{H}= \dfrac{\vec{q}^{\,\textrm{T}}\vec{q}}{2\mathbf{C}}
	+\dfrac{\vec{\phi}^{\,\textrm{T}}\vec{\phi}}{2\mathbf{L}}.
		\label{Eq::H}
\end{equation}
The inverse inductance matrix is defined as
\begin{equation}
	\dfrac{1}{\mathbf{L}}=
	\begin{bmatrix*}[c]
		1/\widetilde{L}_1 & -1/L_{12} \\
		-1/L_{12}         & 1/\widetilde{L}_2
	\end{bmatrix*}
	,
	\label{Eq::Inverse:L:Matrix}
\end{equation}
where~$\widetilde{L}_1=L_1L_{12}/(L_1+L_{12})$ and $\widetilde{L}_2=L_2L_{12}/(L_2+L_{12})$.

The nonconservative elements of the circuit are described by the Rayleigh dissipation function~$\mathcal{D}$~\cite{Mariantoni:2021}. The structure of this function is similar to that of the inductive terms in the Hamiltonian of Eq.~(\ref{Eq::H}), with~$\vec{\phi}\rightarrow\vec{\accentset{\bullet}{\phi}}$ and $\mathbf{L}\rightarrow\mathbf{R}$:
\begin{equation}
	\mathcal{D}=\dfrac{\vec{\accentset{\bullet}{\phi}}^{\,\textrm{T}}
	                   \vec{\accentset{\bullet}{\phi}}}{2\mathbf{R}}.
\end{equation}
The inverse resistance matrix is defined as
\begin{equation}
	\dfrac{1}{\mathbf{R}}=
	\begin{bmatrix*}[c]
		1/\widetilde{R}_1 & -1/R_{12} \\
		-1/R_{12}         & 1/\widetilde{R}_2
	\end{bmatrix*}
	,
	\label{Eq::Inverse:R:Matrix}
\end{equation}
where~$\widetilde{R}_1=R_1R_{12}/(R_1+R_{12})$ and $\widetilde{R}_2=R_2R_{12}/(R_2+R_{12})$.

\subsubsection{Kirchhoff-Heisenberg Circuit Equations}
	\label{Subsubsec::Kirchhoff-Heisenberg:Circuit:Equations}

Following the derivation in App.~\ref{App::Poisson-Rayleigh:Brackets}, which employs the Poisson-Rayleigh brackets, the classical~EOMs of the circuit can be written in the form of Hamilton's equations with damping, the Kirchhoff-Heisenberg equations, which read as
\begin{widetext}
\begin{subequations}
	\begin{empheq}{align}
		\accentset{\bullet}{q}_1
		&=\{(\mathcal{H},\mathcal{D}),(q_1,\accentset{\bullet}{q}_1)\}
		 =-\dfrac{\phi_1}{\widetilde{L}_1}+\dfrac{\phi_2}{L_{12}}
		  -\dfrac{\accentset{\bullet}{\phi}_1}{\widetilde{R}_1}
		  +\dfrac{\accentset{\bullet}{\phi}_2}{R_{12}}
		  \nonumber\\
		&=\{\mathcal{H},q_1\}
		  -\dfrac{\{\mathcal{H},\phi_1\}}{\widetilde{R}_1}
		  +\dfrac{\{\mathcal{H},\phi_2\}}{R_{12}}
		  	\label{Subeq::q1dot}
		  \\[2mm]
		\accentset{\bullet}{q}_2
		&=\{(\mathcal{H},\mathcal{D}),(q_2,\accentset{\bullet}{q}_2)\}
		 = \dfrac{\phi_1}{L_{12}}-\dfrac{\phi_2}{\widetilde{L}_2}
		  +\dfrac{\accentset{\bullet}{\phi}_1}{R_{12}}
		  -\dfrac{\accentset{\bullet}{\phi}_2}{\widetilde{R}_2}
		  \nonumber\\
		&=\{\mathcal{H},q_2\}
		  +\dfrac{\{\mathcal{H},\phi_1\}}{R_{12}}
		  -\dfrac{\{\mathcal{H},\phi_2\}}{\widetilde{R}_2}
		  	\label{Subeq::q2dot}
		  \\[2mm]
		\accentset{\bullet}{\phi}_1
		&=\{(\mathcal{H},\mathcal{D}),(\phi_1,\accentset{\bullet}{\phi}_1)\}
		 = \dfrac{\widetilde{C}_2}{\textrm{det}\mathbf{C}}q_1
		  +\dfrac{C_{12}}{\textrm{det}\mathbf{C}}q_2
		  \nonumber\\
		&=\{\mathcal{H},\phi_1\}
			\label{Subeq::phi1dot}
		 \\[2mm]
		\accentset{\bullet}{\phi}_2
		&=\{(\mathcal{H},\mathcal{D}),(\phi_2,\accentset{\bullet}{\phi}_2)\}
		 = \dfrac{C_{12}}{\textrm{det}\mathbf{C}}q_1
		  +\dfrac{\widetilde{C}_1}{\textrm{det}\mathbf{C}}q_2
		  \nonumber\\
		&=\{\mathcal{H},\phi_2\}.
			\label{Subeq::phi2dot}
	\end{empheq}
\end{subequations}
\end{widetext}

The quantum-mechanical~EOMs are obtained by ``promoting'' the Poisson brackets to commutators. These equations read as
\begin{widetext}
	\begin{subequations}
		\begin{empheq}{align}
			\accentset{\bullet}{\hat{q}}_1
			&= \dfrac{i}{\hbar}[\widehat{\mathcal{H}},\hat{q}_1]
			  -\dfrac{i}{\hbar}\dfrac{[\widehat{\mathcal{H}},\hat{\phi}_1]}{\widetilde{R}_1}
			  +\dfrac{i}{\hbar}\dfrac{[\widehat{\mathcal{H}},\hat{\phi}_2]}{R_{12}}
			  \nonumber\\
			&=-\dfrac{\hat{\phi}_1}{\widetilde{L}_1}
			  +\dfrac{\hat{\phi}_2}{L_{12}}
			  -\dfrac{\widetilde{C}_2}{\widetilde{R}_1\textrm{det}\mathbf{C}}\hat{q}_1
			  -\dfrac{C_{12}}{\widetilde{R}_1\textrm{det}\mathbf{C}}\hat{q}_2
			  +\dfrac{C_{12}}{R_{12}\textrm{det}\mathbf{C}}\hat{q}_1
			  +\dfrac{\widetilde{C}_1}{R_{12}\textrm{det}\mathbf{C}}\hat{q}_2
			  	\label{Subeq::q1hatdot}
			  \\[2mm]
			\accentset{\bullet}{\hat{q}}_2
			&= \dfrac{i}{\hbar}[\widehat{\mathcal{H}},\hat{q}_2]
			  +\dfrac{i}{\hbar}\dfrac{[\widehat{\mathcal{H}},\hat{\phi}_1]}{R_{12}}
			  -\dfrac{i}{\hbar}\dfrac{[\widehat{\mathcal{H}},\hat{\phi}_2]}{\widetilde{R}_2}
			  \nonumber\\
			&= \dfrac{\hat{\phi}_1}{L_{12}}
			  -\dfrac{\hat{\phi}_2}{\widetilde{L}_2}
			  +\dfrac{\widetilde{C}_2}{R_{12}\textrm{det}\mathbf{C}}\hat{q}_1
			  +\dfrac{C_{12}}{R_{12}\textrm{det}\mathbf{C}}\hat{q}_2
			  -\dfrac{C_{12}}{\widetilde{R}_2\textrm{det}\mathbf{C}}\hat{q}_1
			  -\dfrac{\widetilde{C}_1}{\widetilde{R}_2\textrm{det}\mathbf{C}}\hat{q}_2
			  	\label{Subeq::q2hatdot}
			  \\[2mm]
			\accentset{\bullet}{\hat{\phi}}_1
			&= \dfrac{i}{\hbar}[\widehat{\mathcal{H}},\hat{\phi}_1]
			\nonumber\\
			&= \dfrac{\widetilde{C}_2}{\textrm{det}\mathbf{C}}\hat{q}_1
			  +\dfrac{C_{12}}{\textrm{det}\mathbf{C}}\hat{q}_2
			  	\label{Subeq::phi1hatdot}
			  \\[2mm]
			\accentset{\bullet}{\hat{\phi}}_2
			&= \dfrac{i}{\hbar}[\widehat{\mathcal{H}},\hat{\phi}_2]
			\nonumber\\
			&= \dfrac{C_{12}}{\textrm{det}\mathbf{C}}\hat{q}_1
			  +\dfrac{\widetilde{C}_1}{\textrm{det}\mathbf{C}}\hat{q}_2,
			  	\label{Subeq::phi2hatdot}
		\end{empheq}
	\end{subequations}
\end{widetext}
where we substitute Eqs.~(\ref{Subeq::phi1hatdot}) and (\ref{Subeq::phi2hatdot}) into Eqs.~(\ref{Subeq::q1hatdot}) and (\ref{Subeq::q2hatdot}), respectively.

It is worth noting that the quantum-mechanical~EOMs are exactly the same as the classical Eqs.~(\ref{Subeq::q1dot})-(\ref{Subeq::phi2dot}), with the exception that the classical canonical coordinates are substituted by the corresponding quantized coordinates.

For a linear circuit, the quantized coordinates can be written in terms of the bosonic creation and annihilation operators of harmonic oscillators. In the Heisenberg picture, these are time-dependent operators; thus, the coordinates read as:
\begin{subequations}
	\begin{empheq}{align}
		\hat{q}_k(t)
		&=Q_{k0}\left[\dfrac{\hat{o}(t)-\hat{o}^{\dagger}(t)}{i}\right]
			\label{Subeq::qkhat}
			\\[2mm]
		\hat{\phi}_k(t)
		 &=\Phi_{k0}\left[\hat{o}(t)+\hat{o}^{\dagger}(t)\right].
		 	\label{Subeq::phikhat}
	\end{empheq}
\end{subequations}
Here, $\hat{o}=\hat{a}$ for resonator~$1$ and $\hat{b}$ for resonator~2; $Q_{k0}=\sqrt{\hbar/2\widetilde{Z}_k}$ and $\Phi_{k0}=\widetilde{Z}_k Q_{k0}$ are the charge and flux zero-point fluctuations of the $k$-th resonator, and $\widetilde{Z}_k=\sqrt{\widetilde{L}_k/\widetilde{C}_k}$ is the characteristic impedance modified by the presence of the coupling terms. At time~$t_0$, $\hat{o}$ has the usual matrix form with superdiagonal entries~$o_{n,n+1}=\sqrt{n}$ and zero elsewhere; $n=1,2,\ldots$ is the photon number of a harmonic oscillator.

\subsection{Extension to Transmons}
	\label{Subsec::Extension:to:Transmons}

The circuit shown in the diagram of Fig.~\ref{Fig::figure01_res-coupl-circ-diagr_mt_mariantoni} can be easily adjusted to study quantum synchronization in nonlinear circuits, such as transmons. The linear resonators~(1) and (2) can be modified to become transmons by substituting the linear inductors~$L_k$ with Josephson tunnel junctions.

Using the Maclaurin series of the cosine function, the properties of commutators, and the commutation relation between canonical coordinates~$[\hat{\phi},\hat{q}]=i\hbar$, it can be shown that the energy contribution to the circuit Hamiltonian due to the Josephson junction commutes with the charge operator, resulting in
\begin{equation}
-E_{\text{J}0}\dfrac{i}{\hbar}\left[\cos\left(k_{\text{J}}\hat{\phi}\right),\hat{q}\right]=-I_{\text{c}0}\sin{(k_{\text{J}}\hat{\phi})}.
	\label{Eq::EJ:hatq:commutator}
\end{equation}
Here, $E_{\text{J}0}=I_{\text{c}0}/k_{\text{J}}$, $I_{\text{c}0}$ is the critical current of the junction, and $k_{\text{J}}=2\pi/\Phi_0$ is the nonnormalized Josephson constant; $\Phi_0=h/2e$ is the superconducting magnetic flux quantum ($e$ is the electron charge). The expression on the right-hand side of Eq.~(\ref{Eq::EJ:hatq:commutator}) corresponds to the Josephson current. Notably, the very same result is obtained when considering the classical version of Eq.~(\ref{Eq::EJ:hatq:commutator}), where, instead of commutators, the Poisson brackets effectively correspond to a first derivative of the Josephson energy with respect to~$\phi$.

Assuming two lossless transmons coupled only by means of a resistor~$R_{12}$, the circuit's~EOMs read as
\begin{subequations}
	\begin{empheq}{align}
		\accentset{\bullet}{\hat{q}}_k
		&=-I_{\text{c}0}\sin{(k_{\text{J}}\hat{\phi_k})}
		  +\dfrac{(-1)^k}{R_{12}}
		   \left(\dfrac{\hat{q}_1}{C_1}-\dfrac{\hat{q}_2}{C_2}\right)
		\label{Subeq::qkhatdot:transm}
		\\[2mm]
		\accentset{\bullet}{\hat{\phi}}_k
		&= \dfrac{\hat{q}_k}{C_k}.
		\label{Subeq::phikhatdot:transm}
	\end{empheq}
\end{subequations}
It is worth noting that, by linearizing the first term in Eq.~(\ref{Subeq::qkhatdot:transm}), we recover the typical expression for the inductive current in a linear circuit, $-I_{\text{c}0}\sin{(k_{\text{J}}\hat{\phi}_k)}\simeq-\hat{\phi}/L_{\text{J}}$, where~$L_{\text{J}}=1/I_{\text{c}0}k_{\text{J}}$.

Following the notation in the work of Ref.~\cite{Koch:2007} and assuming transmons with identical capacitance~$C_1=C_2=C_{\text{q}}$ and identical critical current, each transmon is characterized by the same inductive energy~$E_{\text{J}0}$ and capacitive energy~$E_{\text{c}}=e^2/2C_{\text{q}}$. By defining the ratio~$\mathcal{R}=E_{\text{J}0}/E_{\text{c}}$, the so-called ``transmon regime'' is realized when~$\mathcal{R}\sim100$. This regime corresponds to a weakly nonlinear scenario, which allows us to use the bosonic operators introduced at the end of the previous section in a perturbative fashion~\cite{Yurke:2004}.

When operating within the transmon regime, it is possible to show that~$\sqrt{8 E_{\text{J}0}E_{\text{c}}}=hf_{\text{ge}}$, where~$f_{\text{ge}}$ is the transition frequency between the ground~$\ket{\text{g}}$ and first excited~$\ket{\text{e}}$ state of the transmon (which effectively behaves as an anharmonic oscillator). From the definition of~$E_{\text{c}}$, it then follows that~$C_{\text{q}}=e^2\sqrt{2\mathcal{R}}/hf_{\text{ge}}$. This expression conveniently allows us to obtain the transmon's capacitance by setting the~$\ket{\text{g}}-\ket{\text{e}}$ transition frequency and~$\mathcal{R}$.

The knowledge of~$L_{\text{J}}$ and $C_{\text{q}}$ makes it possible to calculate the linearized impedance of a transmon, which reads as~$Z_{\text{q}}=\sqrt{L_{\text{J}}/C_{\text{q}}}$. From the impedance, it is finally possible to find the zero-point fluctuations for the charge and flux in the transmon by substituting~$\widetilde{Z}_k$ with~$Z_{\text{q}}$ in the expressions after Eq.~(\ref{Subeq::phikhat}).

\subsection{Nonconservative Pathological Circuit}
	\label{Subsec::Nonconservative:Pathological:Circuit}

\begin{figure}[t!]
	\centering
	\includegraphics[width=1.0\columnwidth]{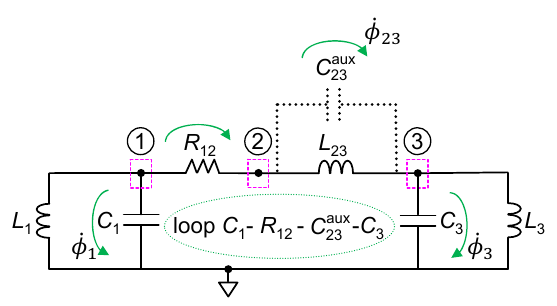}
	\caption{Circuit diagram of a pathological circuit, where two lossless~$LC$ resonators are coupled through a nonconservative~$RL$-series circuit. Conventions and signs are the same as in Fig.~\ref{Fig::figure01_res-coupl-circ-diagr_mt_mariantoni}. The dotted black lines indicate the auxiliary capacitor~$C_{23}^{\text{aux}}$. In this case, there are three independent nodes ($\Circled{1}$, $\Circled{2}$, and $\Circled{3}$), corresponding to three~KCLs and, thus, three~EOMs (and DOFs).}
		\label{Fig::figure02_pathol-circ-diagr_mt_mariantoni_v1-00}
\end{figure}

The circuit depicted in Fig.~\ref{Fig::figure01_res-coupl-circ-diagr_mt_mariantoni} operates under the assumption of an ideal resistor with resistance~$R_{12}$. However, in practical applications, this ideal resistor is often accompanied by spurious elements, such as a \emph{parasitic series inductor} with inductance~$L_{23}$. The circuit diagram in Fig.~\ref{Fig::figure02_pathol-circ-diagr_mt_mariantoni_v1-00} showcases a more realistic device configuration, incorporating the presence of~$L_{23}$. For the purpose of focusing on the impact of~$L_{23}$, we intentionally exclude any local resistors~$R_k$ and reactive coupling elements~$L_{12}$ and $C_{12}$.

The physical origin of~$L_{23}$ can be understood through the examination of a real resistor operating at low temperatures (e.g., at~$T\sim\SI{10}{\milli\kelvin}$), often constructed using a strip made from a normal conducting metal. The resistance is directly proportional to the length of the strip, as does its inductance. Interestingly, even if it were feasible to develop an ideal point-like resistor, employing a strip for spacing between two resonators offers the advantage of isolating the resonators and reducing crosstalk~\footnote{The amount of spacing can be adjusted by introducing meanders in the strip.}. In Sec.~\ref{Subsec::RESULTS:Pathological:Circuit}, we provide a brief analysis of the realistic parameters associated with the resistance and inductance of a conducting strip. Notably, the inclusion of the inductance~$L_{23}$ has significant implications in circuit theory, extending beyond its role as a parasitic element.

The electrical circuit depicted in Fig.~\ref{Fig::figure02_pathol-circ-diagr_mt_mariantoni_v1-00} is characterized by~$N=4$ nodes and $B=6$ branches. Notably, in this case, the relationship~$N-1=3=B-N+1$ holds true. Consequently, constraints may be imposed by either applying~KVLs or KCLs. However, the use of~KVLs proves to be more advantageous. This preference is justified by the potential inclusion of additional local resistors in parallel with~$C_1$ and $C_3$, resulting in a modified topology where the number of independent nodes remains the same, $N-1=3$, but~$B-N+1=5$. By following the methodology delineated in Sec.~\ref{Subsec::Circuit:Theory:of:Two:Resonators:Interacting:Via:a:Nonconservative:Coupler}, we can derive the circuit's Hamiltonian, which reads as
\begin{eqnarray}
	\widehat{\mathcal{H}}_{\text{inc}}
	&=&\sum_{k=1,3}\left(\dfrac{\hat{q}_k^2}{2 C_k}+\dfrac{\hat{\phi}_k^2}{2 L_k}\right)
	\nonumber\\
	&+&\dfrac{\hat{\phi}_{23}^2}{2 L_{23}},
	\label{Eq::Hhat:inc}
\end{eqnarray}
where~$\hat{\phi}_{23}$ is the quantized flux on~$L_{23}$.

The Hamiltonian~$\widehat{\mathcal{H}}_{\text{inc}}$ is incomplete because it does not include a term associated with the conjugate variable of~$\hat{\phi}_{23}$, $\hat{q}_{23}$, which is missing. Upon deriving the~EOMs associated with~$\widehat{\mathcal{H}}_{\text{inc}}$, it is evident that~$\accentset{\bullet}{\hat{\phi}}_{23}=0$, leading to the solution~$\hat{\phi}_{23}(t)=const.$ for all values of~$t$. However, a simple analysis based on Kirchhoff's circuit laws indicates that this solution is generally incorrect~\footnote{This is especially noticeable when~$L_{23} \gg L_1,L_3$.}. The circuit illustrated in Fig.~\ref{Fig::figure02_pathol-circ-diagr_mt_mariantoni_v1-00} exemplifies a peculiar case, deemed as a~\emph{pathological circuit}. While Kirchhoff's circuit laws offer a straightforward method for solving this type of circuit, the application of the Hamiltonian formalism proves to be inadequate for achieving a solution in this specific case.

To rectify this inconsistency, an additional circuit element corresponding to~$\hat{q}_{23}$ must be introduced. This supplementary component is identified as an~\emph{auxiliary capacitor} denoted as~$C_{23}^{\text{aux}}$. This capacitor must be placed in parallel with~$L_{23}$ in order to maintain the established count of independent nodes~$N-1=3$, thereby preserving the system's~DOFs at three. The parallel connection of~$L_{23}$ and $C_{23}^{\text{aux}}$ results in the auxiliary resonator~$23$. This auxiliary resonator allows us to define a complete set of three (quantized) conjugate variables, $\{\hat{\phi}_k,\hat{q}_k\}$, where now~$k=1,23,3$. The Hamiltonian of the complete circuit, including~$C_{23}^{\text{aux}}$, reads as
\begin{equation}
	\widehat{\mathcal{H}}_{\text{compl}}=\widehat{\mathcal{H}}_{\text{inc}}+\dfrac{\hat{q}_{23}^2}{2 C_{23}^{\text{aux}}}.
		\label{Eq::Hhat:compl}
\end{equation}

The auxiliary capacitor~$C_{23}^{\text{aux}}$ could be seen as the parasitic element of a nonideal inductor. However, in realistic inductor's models, this parasitic element is typically connected in parallel with the series of~$R_{12}$ and $L_{23}$, adding complexity to the analysis. It is more beneficial to view~$C_{23}^{\text{aux}}$ as an \textit{ad hoc} component, included solely to complete the circuit's Hamiltonian. This perspective allows us to treat~$C_{23}^{\text{aux}}$ as a virtual element that can be ``inserted'' into the physical circuit as necessary to resolve a pathology.

The dissipation function associated with the only nonconservatiove element of the circuit, $R_{12}$, reads as
\begin{equation}
	\widehat{\mathcal{D}}_{12}=\dfrac{\left(\accentset{\bullet}{\hat{\phi}}_1-\accentset{\bullet}{\hat{\phi}}_{23}-\accentset{\bullet}{\hat{\phi}}_3\right)^2}{2R_{12}}.
		\label{Eq::Dhat:23}
\end{equation}
The voltage on~$R_{12}$ is found from KVL around the loop~$C_1$-$R_{12}$-$C_{23}^{\text{aux}}$-$C_3$ (see Fig.~\ref{Fig::figure02_pathol-circ-diagr_mt_mariantoni_v1-00}).

The quantized~EOMs are obtained from~$\widehat{\mathcal{H}}_{\text{compl}}$ and~$\widehat{\mathcal{D}}_{12}$ using the Poisson-Rayleigh parenthesis and read as
\begin{subequations}
	\begin{empheq}{align}
		\accentset{\bullet}{\hat{q}}_k
		&=-\dfrac{\hat{\phi}_k}{L_k}
		+(-1)^u\dfrac{1}{R_{12}}\left[\sum_{\ell=1,23,3}(-1)^v\dfrac{\hat{q}_{\ell}}{C_{\ell}}\right]
			\label{Subeq::qkhatdot:path}
		\\[2mm]
		\accentset{\bullet}{\hat{\phi}}_k
		&= \dfrac{\hat{q}_k}{C_k},
			\label{Subeq::phikhatdot:path}
	\end{empheq}
\end{subequations}
where~$C_{23}$ must be intended as~$C_{23}^{\text{aux}}$ and the indexes are paired as~$(k,u)\in\{(1,1),(23,2),(3,2)\}$ and $(\ell,v)\in\{(1,2),(23,1),(3,1)\}$. It should be pointed out that in Eq.~(\ref{Subeq::qkhatdot:path}), we express the dissipative current on~$R_{12}$ using the charges~$\hat{q}_{\ell}$ rather than the voltages~$\accentset{\bullet}{\hat{\phi}}_{\ell}$. This approach results in the introduction of the time constants~$R_{12}C_{\ell}$.

The auxiliary element~$C_{23}^{\text{aux}}$ is essential for completing the circuit's Hamiltonian and deriving the correct~EOMs. However, this element is not present in the original physical circuit. As a result, it is necessary to ultimately eliminate its influence from the dynamics of the circuit. This objective is accomplished by approaching the limiting scenario where~$C_{23}^{\text{aux}}\to0^+$. In this limit, $C_{23}^{\text{aux}}$ effectively transforms into an open~circuit, exhibiting negligible influence.

This condition leads to an infinite resonance frequency for resonator~$23$, $f_{23\text{r}}\to+\infty$. In the quantized case, the discrete energy levels of this resonator exhibit an infinitely large separation. As a result, resonator~$23$ remains consistently in the energy ground state and can be seen as analogous to a classical coupling inductor~$L_{23}$.

In practice (e.g., in numerical simulations), implementing the condition~$C_{23}^{\text{aux}}\to0^+$ involves setting the value of~$C_{23}^{\text{aux}}$ to be at least two orders of magnitude smaller than the next smallest capacitor in the circuit. This approach simulates well the ideal open-circuit condition, as we show in Sec.~\ref{Subsec::RESULTS:Pathological:Circuit}.

\subsection{Numerical Simulations}
	\label{Subsec::Numerical:Simulations}

Linear systems, such as the one represented by Eqs.~(\ref{Subeq::q1hatdot})-(\ref{Subeq::phi2hatdot}), can be conveniently solved using analytical methods. However, the situation presents greater challenges when dealing with nonlinear systems (e.g., in presence of transmon circuits), since analytical solutions are not readily attainable. Therefore, to make comparisons, we elect to solve all the systems in this study through the use of numerical simulations.

In the simulations, we normalize the quantized conjugate variables by dividing them by their respective zero-point fluctuations. In the EOMs, we express the time derivative of these variables as~$\accentset{\bullet}{\hat{q}}_k/Q_{k0}$ and $\accentset{\bullet}{\hat{\phi}}_k/\Phi_{k0}$. This normalization yields prefactors in the form of characteristic impedances, transimpedances, or dimensionless normalization factors on the right-hand side of the~EOMs. For instance, in Eq.~(\ref{Subeq::q1hatdot}), the normalization gives rise to the impedance~$\Phi_{10}/Q_{10}$, the transimpedance~$\Phi_{20}/Q_{10}$, and the normalization factor~$Q_{20}/Q_{10}$.

The initial conditions for each quantized variable are obtained from the entries of the normalized matrices of Eqs.~(\ref{Subeq::qkhat}) and (\ref{Subeq::phikhat}), namely~$\hat{q}_k(t_0)/Q_{k0}$ and $\hat{\phi}_k(t_0)/\Phi_{k0}$. To simplify the numerical complexity, we restrict the Hilbert space associated with each pair of conjugate variables~$\hat{q}_k,\hat{\phi}_k$ (i.e., with each DOF) to a dimension of~$n=3$ for the simple linear circuit in Fig.~\ref{Fig::figure01_res-coupl-circ-diagr_mt_mariantoni}, $n=4$ for the nonlinear circuit with transmons, and $n=2$ for the pathological circuit in Fig.~\ref{Fig::figure02_quant-charges_mt_mariantoni_v1-04}. It is important to stress that we only consider scenarios with quantized circuits prepared in the single excitation subspace, in presence of damping and absence of any driving. This simplification, which allows us to focus on the circuit aspect of our study rather than numerical technicalities, makes it possible not to incur in truncation mistakes, while maintaining a manageable size of the Hilbert space. In our simulations, increasing the size of the Hilbert space would not lead to more accurate results.

The time evolution of the expectation value for the charge and flux observables is computed by assuming an initial state~$\ket{\Psi_k(t_0)}$ for each DOF,
\begin{subequations}
	\begin{empheq}{align}
		\braket{\hat{q}_k}
		&=\braket{\Psi_k(t_0)|\hat{q}_k(t)|\Psi_k(t_0)}
		\label{Subeq::expval:qkhat}
		\\[2mm]
		\braket{\hat{\phi}_k}
		&=\braket{\Psi_k(t_0)|\hat{\phi}_k(t)|\Psi_k(t_0)}.
		\label{Subeq::expval:phikhat}
	\end{empheq}
\end{subequations}
In all simulations, we assume~$\Psi_k(t_0)=(\alpha_k\ket{0}+\beta_k\ket{1})/\sqrt{\alpha_k\alpha_k^{\ast}+\beta_k\beta_k^{\ast}}$ (i.e., state~$\ket{2}$ is always unpopulated at~$t_0$).

\section{RESULTS}
	\label{Sec::RESULTS}

In Sec.~\ref{Subsec::RESULTS:Simple:Circuit}, we present the numerical results on the quantum synchronization for a simple circuit with resistively-coupled resonators; we consider different regimes. In Sec.~\ref{Subsec::RESULTS:Transmons}, we show the results for a circuit with transmons. Finally, in Sec.~\ref{Subsec::RESULTS:Pathological:Circuit}, we present our findings for the case of a pathological circuit. We study different regimes, with the aim of understanding the role played by an auxiliary circuit element.

\subsection{Simple Circuit}
	\label{Subsec::RESULTS:Simple:Circuit}

We analyze the two-resonator circuit shown in Fig.~\ref{Fig::figure01_res-coupl-circ-diagr_mt_mariantoni} by solving numerically the linear system of Eqs.~(\ref{Subeq::q1hatdot})-(\ref{Subeq::phi2hatdot}) (see Sec.~\ref{Subsubsec::Kirchhoff-Heisenberg:Circuit:Equations}). Our study focuses on two different operating regimes: (1) Resonant and low local loss. (2) Detuned and high local loss. The simulation parameters are reported in Table~\ref{Table::Simulation:Parameters}. In both regimes, we assume that resonator~$1$ is initially prepared in an equal superposition state with~\mbox{$\alpha_1=\beta_1=1$}, while resonator~$2$ is in the vacuum state with~$\alpha_2=1$ and $\beta_2=0$.

\begin{table}[b!]
	\caption{Simulation parameters. We elect to set the inductance and resonance frequency of the~$k$-th resonator, $L_k$ and $f_{k\textrm{r}}$. The corresponding capacitance can therefore be found from~$C_k=1/(4\pi^2f_{k\textrm{r}}^2L_k)$.
		\label{Table::Simulation:Parameters}}
	\begin{center}
		\begin{ruledtabular}
			\begin{tabular}{lrr}
				Parameter & Regime~(1) & Regime~(2) \\
				\hline
				\raisebox{0mm}[4mm][0mm]{$L_1$} (\si{\nano\henry}) & $1$ & $1$ \\
				\raisebox{0mm}[3mm][0mm]{$f_{1\text{r}}$} (\si{\mega\hertz}) & $5000$ & $5000$ \\
				\raisebox{0mm}[3mm][0mm]{$C_1$} (\si{\pico\farad}) & $1.01$ & $1.01$ \\
				\raisebox{0mm}[4mm][0mm]{$R_1$} (\si{\mega\ohm}) & $15.71$ & $0.1571$ \\
				\raisebox{0mm}[3mm][0mm]{$R_{12}$} (\si{\kilo\ohm}) & $4$ & $4$ \\
				\raisebox{0mm}[3mm][0mm]{$L_{12}=10 L_1$} (\si{\nano\henry}) & $10$ & $10$ \\
				\raisebox{0mm}[3mm][0mm]{$C_{12}=C_1/50$} (\si{\femto\farad}) & $20.26$ & $20.26$ \\
				\raisebox{0mm}[3mm][0mm]{$L_2$} (\si{\nano\henry}) & $1$ & $1$ \\
				\raisebox{0mm}[3mm][0mm]{$f_{2\text{r}}$} (\si{\mega\hertz}) & $5000$ & $5050$ \\
				\raisebox{0mm}[3mm][0mm]{$C_2$} (\si{\pico\farad}) & $1.01$ & $0.99$ \\
				\raisebox{0mm}[3mm][0mm]{$R_2$} (\si{\mega\ohm}) & $15.71$ & $0.1571$ \\
			\end{tabular}
		\end{ruledtabular}
	\end{center}
\end{table}

\begin{figure*}[t!]
	\centering
	\includegraphics[width=1.0\textwidth,trim={19cm 0 19cm 0},clip]{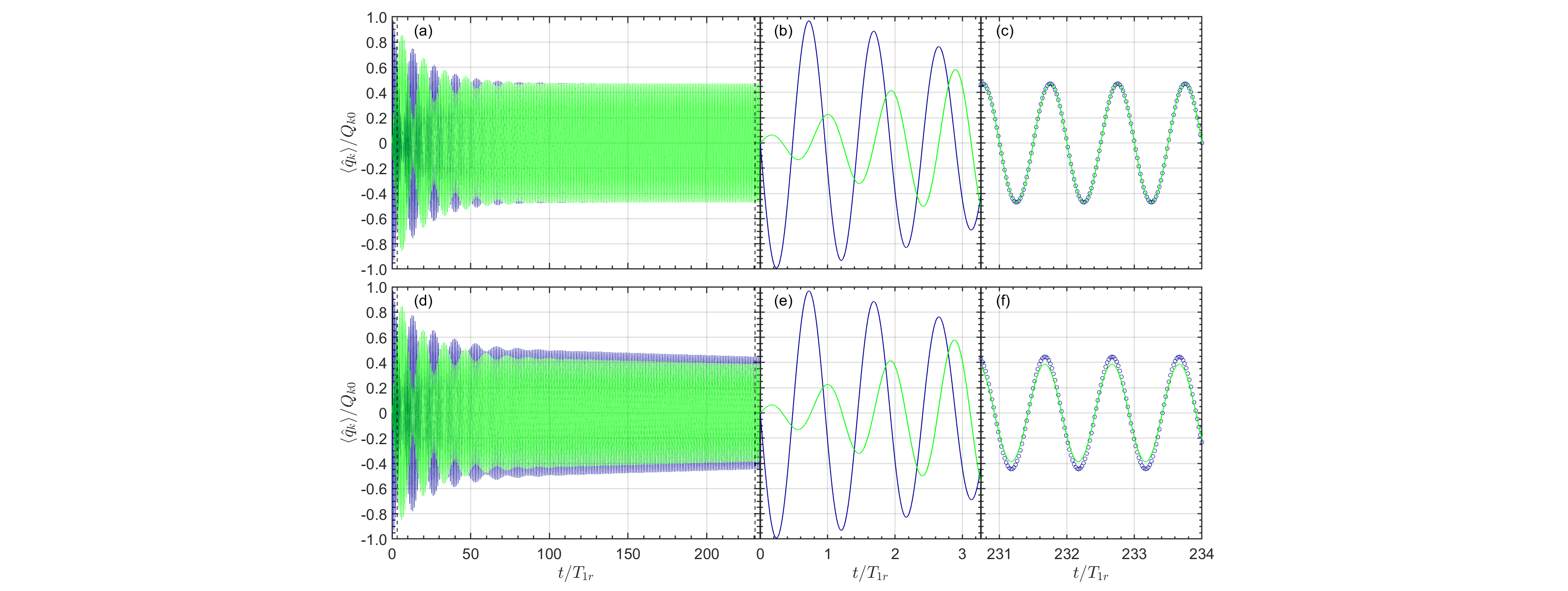}
	\caption{Exploring quantum synchronization between two resonators connected by a nonconservative coupler, illustrated in the circuit diagram of Fig.~\ref{Fig::figure01_res-coupl-circ-diagr_mt_mariantoni}. The temporal evolution of the normalized expectation value for the quantized charges~$\braket{\hat{q}_k}/Q_{k0}$ is plotted with respect to normalized time~$t/T_{1\text{r}}$, where the normalization period is given by~$T_{1\text{r}}=1/f_{1\text{r}}$ (we use the normalization period~$T_{1\text{r}}$ of resonator~$1$ for both resonators). (a)-(c): regime~(1). (d)-(f): regime~(2). The dark blue~(dark gray) lines correspond to~$\braket{\hat{q}_1}/Q_{10}$, and the light green~(light gray) lines to~$\braket{\hat{q}_2}/Q_{20}$. (b) and (e): focusing on the initial periods [prior to the vertical dashed black lines in~(a) and (d)], where the oscillations exhibit complete phase mismatch. (c) and (f): showcasing a few post-synchronization periods [following the vertical dashed black lines in~(a) and (d)]; the data points pertaining to~$\braket{\hat{q}_1}/Q_{10}$ are marked by open dark blue~(dark gray) circles for clarity. Notably in~(f), there is an evident asymmetry in peak amplitudes due to detuning.}
	\label{Fig::figure02_quant-charges_mt_mariantoni_v1-04}
\end{figure*}

\begin{figure}[t!]
	\centering
	\includegraphics[width=1.0\columnwidth,trim={23cm 1cm 36.5cm 1cm},clip]{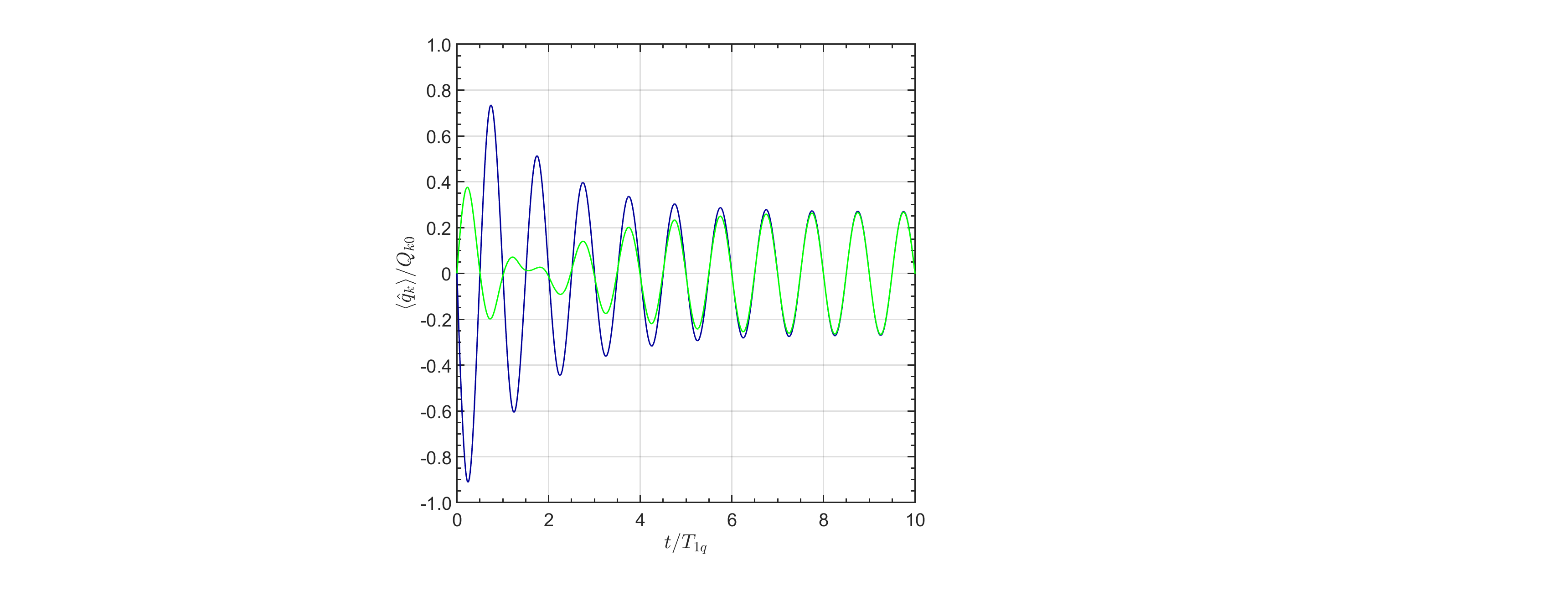}
	\caption{Quantum synchronization between two identical, lossless transmons coupled by a resistor, as explained in Sec.~\ref{Subsec::Extension:to:Transmons}. The temporal evolution of the normalized expectation value for the quantized charges~$\braket{\hat{q}_k}/Q_{k0}$ is plotted with respect to normalized time~$t/T_{\text{ge}}$, where the normalization period is given by~$T_{\text{ge}}=1/f_{\text{ge}}$. The dark blue~(dark gray) lines correspond to~$\braket{\hat{q}_1}/Q_{10}$, and the light green~(light gray) lines to~$\braket{\hat{q}_2}/Q_{20}$ (in this case, $Q_{10}=Q_{20}$). Quantum synchronization is reached within less than ten periods.}
		\label{Fig::figure04_transmons-circ-sims_mt_mariantoni}
\end{figure}

\begin{figure*}[t!]
	\centering
	\includegraphics[width=1.0\textwidth,trim={13.1cm 0 17.5cm 0},clip]{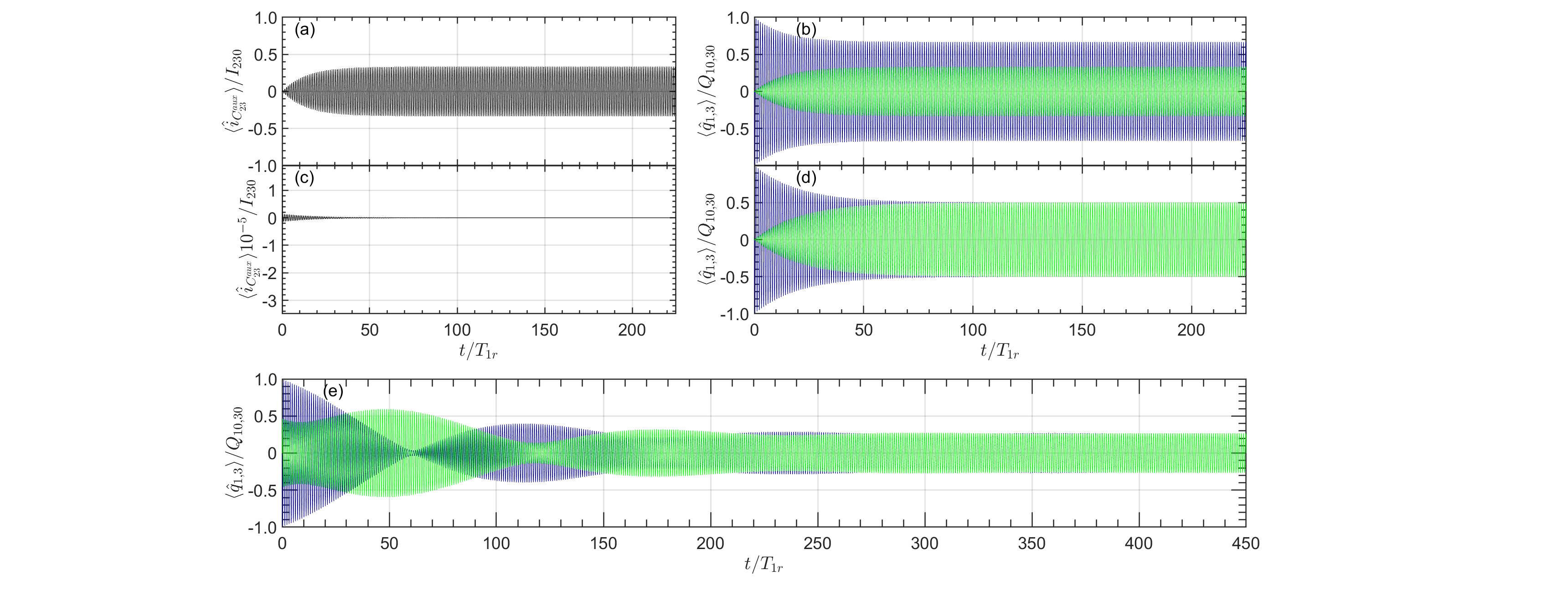}
	\caption{Quantum synchronization for the nonconservative pathological circuit illustrated in Fig.~\ref{Fig::figure02_pathol-circ-diagr_mt_mariantoni_v1-00}. The temporal evolution of the normalized expectation value for the quantized current on~$C_{23}^{\text{aux}}$, $\braket{\hat{i}_{C_{23}^{\text{aux}}}}/I_{230}$~[(a) and (c)], and for the quantized charges on resonators~$1$ and $3$, $\braket{\hat{q}_{1,3}}/Q_{10,30}$~[(b), (d), and (e)], is plotted with respect to normalized time~$t/T_{1\text{r}}$. In this case, $Q_{10,30}=\sqrt{\hbar/2Z_{1,3}}$, with~$Z_{1,3}=\sqrt{L_{1,3}/C_{1,3}}$. (a)-(b): scenario with~$C_{23}^{\text{aux}}=C_1$, $R_{12}=\SI{4}{\kilo\ohm}$, and $L_{23}=L_1$ (other parameters and initial states described in the main text). (c)-(d): scenario with~$C_{23}^{\text{aux}}=C_1/1000$, $R_{12}=\SI{4}{\kilo\ohm}$, and $L_{23}=L_1$. (e) More realistic scenario with~$C_{23}^{\text{aux}}=C_1/1000$, $R_{12}=\SI{1}{\kilo\ohm}$, and $L_{23}=100L_1$. The dark blue~(dark gray) lines correspond to~$\braket{\hat{q}_1}/Q_{10}$, and the light green~(light gray) lines to~$\braket{\hat{q}_3}/Q_{30}$. An evident peak amplitude asymmetry is observed in~(b) attributed to the substantial impact of~$C_{23}^{\text{aux}}$. In~(e), significant asynchronous dynamics occur initially, due to the large inductance~$L_{23}$.}
	\label{Fig::figure05_pathol-circ-sims_mt_mariantoni_v1-00}
\end{figure*}

Figure~\ref{Fig::figure02_quant-charges_mt_mariantoni_v1-04} displays the temporal evolution of the expectation value for the quantized charge in both resonators. By inverting the quantized form of Eq.~(\ref{Eq::C:vecphidot:vecq}), we can obtain the corresponding values for the voltage operators associated with these charges. The behavior of the flux variables is also similar, with their expectation values being~\SI{90}{\degree} out of phase compared to the charge case.

Regime~(1) exemplifies a scenario based on the experimental parameters of typical superconducting quantum devices fabricated using planar technology on silicon substrates. We assume perfect resonance, $f_{1\text{r}}=f_{2\text{r}}=f_{\text{r}}$, and low-loss resonators characterized by an internal quality factor of~$Q_{\textrm{int},k}=\num{5e5}$~\footnote{It should be noted that the chosen quality factor is intentionally lower than the state-of-the-art value of approximately a million~\cite{Earnest:2018}, allowing for experimental flexibility.}. To determine the local resistance of each resonator, we employ the equation~$R_k=Q_{\textrm{int},k}Z_k$, where~$Z_k=\sqrt{L_k/C_k}$ represents the characteristic impedance of an individual resonator without considering any coupling modification.

The results for regime~(1) are shown in Fig.~\ref{Fig::figure02_quant-charges_mt_mariantoni_v1-04}~(a)-(c). As time progresses, $\braket{\hat{q}_1(t)}$ and $\braket{\hat{q}_2(t)}$ become synchronized, meaning they have the same peak amplitude and oscillate in phase with one another at all times. This synchronization happens after a transient period dictated by the time constant~$\tau_{RC}=R_{12}C$, with~$C=C_1=C_2$. After reaching a steady-state condition, the synchronized oscillations occur with a period equal to~$1/f_{\text{r}}$ and their peak amplitude stabilizes at a constant value of~$0.5$ (see discussion in Sec.~\ref{Sec::DISCUSSION}). Moreover, it is worth mentioning that the impact of local loss is negligible, even over long time intervals. Remarkably, for several hundred periods, the temporal behavior remains practically unchanged compared to the lossless scenario.

It can easily be shown that in regime~(1), when~\mbox{$R_{12}\to+\infty$} (essentially behaving as an open circuit), the system exhibits dynamics akin to the Jaynes-Cummings model in the single excitation subspace, or a pair of qubits coupled either capacitively or inductively. It is only with the introduction of finite values of~$R_{12}$ that the system transitions into a state of quantum synchronization.

Regime~(2) delves into a scenario where there is a detuning of~$f_{2\textrm{r}}-f_{1\textrm{r}}=\Delta=\SI{50}{\mega\hertz}$ and significantly high local losses, represented by internal quality factors approximately~$100$ times lower than in regime~(1) (i.e., small values of~$R_k$). This detuning can be intentionally set by the experimenter or may arise as a result of nonidealities during the fabrication process of the device. While such high levels of local losses are not common in standard experiments, they help clearly illustrate the impact of damping on quantum synchronization.

The results for regime~(2) can be seen in Fig.~\ref{Fig::figure02_quant-charges_mt_mariantoni_v1-04}~(d)-(f), where despite detuning and high loss, quantum synchronization is achieved following a transient period similar to regime~(1). It is worth noting that upon reaching a steady-state condition and before significant damping sets in, the amplitude of the synchronized oscillations exceeds~$0.5$ for~$\braket{\hat{q}_1}$ and falls below~$0.5$ for~$\braket{\hat{q}_2}$. This observation underscores the potential of detuning as a method for calibrating, controlling, and turning off synchronization, demonstrating its utility in tunable devices such as transmon circuits. In the presence of high losses, the peak amplitude of the synchronized oscillations decays exponentially over time, albeit at a slow rate. These findings provide reassurance of the synchronization phenomenon's robustness in the presence of device imperfections.

\subsection{Transmons}
	\label{Subsec::RESULTS:Transmons}

We study the two-transmon circuit introduced in Sec.~\ref{Subsec::Extension:to:Transmons} by solving numerically the nonlinear system of Eqs.~(\ref{Subeq::qkhatdot:transm}) and (\ref{Subeq::phikhatdot:transm}). For each transmon, we consider~$f_{\text{ge}}=\SI{5}{\giga\hertz}$ and an anharmonicity ratio~$\mathcal{R}=50$. We set the coupling resistor between the two transmons to be~$R_{12}=\SI{4}{\kilo\ohm}$. We assume that transmon~$1$ is initially prepared in an equal superposition state with~$\alpha_1=\beta_1=1$, while transmon~$2$ is in a state with~$\alpha_2=0.2$ and $\beta_2=-0.8$.

The results of our simulations are displayed in Fig.~\ref{Fig::figure04_transmons-circ-sims_mt_mariantoni}. Even in presence of weak nonlinearities, quantum synchronization is achieved after a short transient time. It is worth noting that, while there is a difference between the curves for the linear and nonlinear systems, such a difference is very small.

\subsection{Pathological Circuit}
	\label{Subsec::RESULTS:Pathological:Circuit}

Finally, we address the~EOMs pertaining to the pathological circuit illustrated in Fig.~\ref{Fig::figure02_pathol-circ-diagr_mt_mariantoni_v1-00}, Eqs.~(\ref{Subeq::qkhatdot:path}) and (\ref{Subeq::phikhatdot:path}) (see Sec.~\ref{Subsec::Nonconservative:Pathological:Circuit}). Two distinct cases are investigated, with the first case mirroring regime~(1) studied in the simpler two-resonator circuit. The parameters used align with those reported in Table~\ref{Table::Simulation:Parameters}. However, we disregard any local resistances and set~$L_{23}=L_1$, $L_3=L_2$, and $C_3=C_2$, alongside choosing an appropriate value for the auxiliary capacitor~$C_{23}^{\text{aux}}$. To assess the impact of this element, comparison is made between simulation outcomes for~$C_{23}^{\text{aux}}=C_1$ and $C_{23}^{\text{aux}}=C_1/1000$, the second choice mimicking an open circuit scenario. The open-circuit configuration enables a direct comparison between the findings of the simple two-resonator circuit and the pathological circuit. For coherence in this comparison, we opt again for the initial states with~$\alpha_1=\beta_1=1$ for resonator~$1$ and with~$\alpha_3=1$ and $\beta_3=0$ for resonator~$3$ (which, in this case, plays the role of resonator~$2$). Furthermore, we assume that the coupling auxiliary resonator~$23$ is initially prepared in the vacuum state with~$\alpha_{23}=1$ and $\beta_{23}=0$.

Figure~\ref{Fig::figure05_pathol-circ-sims_mt_mariantoni_v1-00}~(a) and (c) illustrates the temporal evolution of the expectation value for the quantized current carried by the auxiliary capacitor~$C_{23}^{\text{aux}}$, $\braket{\hat{i}_{C_{23}^{\text{aux}}}}$, under two different conditions. Panel~(a) represents the scenario where~$C_{23}^{\text{aux}}=C_1$, while panel~(c) depicts the case where~$C_{23}^{\text{aux}}=C_1/1000$ (corresponding to an open circuit). The current can be found either by differentiating numerically the charge~$\hat{q}_{23}$, thus obtaining~$\accentset{\bullet}{\hat{q}}_{23}$, or by applying~KCL at node~$\Circled{2}$, that is, as in Eq.~(\ref{Subeq::qkhatdot:path}) for~$k=23$. The mean current is normalized by the zero-point current associated with resonator~$23$, $I_{230}=\sqrt{hf_{23\text{r}}/2L_{23}}$.

In the first scenario ($C_{23}^{\text{aux}}=C_1$), a significant current is observed, owing to the relatively low reactance of the auxiliary capacitor at the resonance frequency~$f_{\text{r}}=\SI{5}{\giga\hertz}$, $X_{C_{23}^{\text{aux}}}\approx\SI{31.52}{\ohm} \approx X_{L_{23}}$. As anticipated, this current is several orders of magnitude smaller and further diminishes over time in the open circuit case, where~$X_{C_{23}^{\text{aux}}}^{\text{OC}}\approx\SI{31.52}{\kilo\ohm} \gg X_{C_{23}^{\text{aux}}}$.

When the value of~$C_{23}^{\text{aux}}$ is large, $C_{23}^{\text{aux}}=C_1$, asynchronous oscillations are observed in the expectation value of the quantized charges for resonators~$1$ and $3$, shown in Fig.~\ref{Fig::figure05_pathol-circ-sims_mt_mariantoni_v1-00}~(b). These oscillations exhibit the same phase but significantly different amplitudes~\footnote{We adhere to the strict definition employed in ac~electric networks, requiring precise alignment of phase, signal amplitude, and frequency for synchronization to be attained.}. Contrarily, perfect synchronization is achieved for very small values of~$C_{23}^{\text{aux}}$, $C_{23}^{\text{aux}}=C_1/1000$, as depicted in Fig.~\ref{Fig::figure05_pathol-circ-sims_mt_mariantoni_v1-00}~(d). In this scenario, the post-synchronization oscillations mirror those in Fig.~\ref{Fig::figure02_quant-charges_mt_mariantoni_v1-04}~(a).

In the second case, a more realistic set of coupling parameters is considered. These revised parameters account for the requirement of a long strip of conductor to realize a high resistance value of~$R_{12}\sim\SI{1}{\kilo\ohm}$. The extended conducting strip inherently introduces a considerable parasitic inductance~$L_{23} \gg L_1$. For example, assume a thin-film gold~(Au) strip with a sheet resistance of~$R_{\text{s}}=\rho_{\text{Au}}/d\approx\SI{3}{\milli\ohm}$, where~$\rho_{\text{Au}}$ denotes the~Au resistivity at~\SI{1}{\kelvin} and $d=\SI{75}{\nano\meter}$ signifies a typical film thickness. For a strip with width~$W=\SI{500}{\nano\meter}$ and length~$\ell_{\text{L}}=\SI{20}{\centi\meter}$, the strip's resistance amounts to about~\SI{1}{\kilo\ohm}, allowing us to set~$R_{12}=\SI{1}{\kilo\ohm}$ in simulations. Consequently, an approximately~$100$-fold increase in parasitic inductance compared to~$L_1$ is estimated, leading to~$L_{23}=100L_1$, which is the value used in these simulations. The remaining parameters echo those of the first case, with~$C_{23}^{\text{aux}}=C_1/1000$. Resonator~$1$ remains in the usual initial state with~$\alpha_1=\beta_1=1$, while resonator~$2$ is initialized in a state with~$\alpha_2=0.2$ and $\beta_2=-0.8$ to explore a different dynamics. The auxiliary resonator~$23$ is initialized again in the vacuum state.

The temporal evolution for the quantized charges in resonators~$1$ and $3$ is shown in Fig.~\ref{Fig::figure02_quant-charges_mt_mariantoni_v1-04}~(e). Despite the significant asynchronous initial dynamics due to the large value of~$L_{23}$, the charges become completely synchronized after a short transient.

\section{DISCUSSION}
	\label{Sec::DISCUSSION}


When studying regime~(1) in Sec.~\ref{Subsec::RESULTS:Simple:Circuit}, the coupling between resonators~$1$ and $2$ is mainly due to~$R_{12}$. In fact, the steady-state impedances associated with~$L_{12}$, $i\omega L_{12}$ ($\omega=2\pi f$), and with $C_{12}$, $1/i\omega C_{12}$, approach that of an open circuit for~$f\simeq f_{k\textrm{r}}$. The results of Fig.~\ref{Fig::figure02_quant-charges_mt_mariantoni_v1-04}~(a) show that, after reaching synchronization, the coordinates persist oscillating without any damping (neglecting local loss). In the quantum case, this is equivalent to stating the post-synchronization system is characterized by unitary dynamics. This statement seems rather counterintuitive considering the coupling coefficient between the two resonators is a \textit{purely lossy element}.


To ease the intuition, we resort to a mechanical analogy of our circuit. The circuit's dynamics can be mapped into the small oscillations of two pendulums coupled by means of a soft, lossy spring. Softness emulates a small inductive (and capacitive) coupling; loss emulates electrical resistance.

After synchronization, the two pendulums oscillate in phase; thus, no energy can excite the spring, which remains displaced at all times at the same distance~$d_0$ as the two unperturbed pendulums. We remind that, in the lossless case, two coupled pendulums are characterized by two normal modes. The parallel mode, where the pendulums oscillate in phase with angular frequency~$\omega_{\textrm{r}}=\omega_{\textrm{par}}=2\pi f_{\textrm{r}}$ (assuming~$\Delta=0$). The symmetric mode, where the pendulums oscillate~\SI{180}{\degree} out of phase with an angular frequency~$\omega_{\textrm{symm}}>\omega_{\textrm{r}}$ due to the coupling. The post-synchronization dynamics of our lossy circuit resembles those of the parallel mode.

To confirm our intuition, we must find the complex eigenfrequencies~$s$ of the circuit and show that, in the steady-state regime, they lead to the expected undamped parallel mode. In order to simplify the analytical calculation, we assume purely lossy coupling with finite~$R_{12}$ and no reactive coupling (i.e., $L_{12}\rightarrow+\infty$ and $C_{12}\rightarrow0^+$). Under these conditions, the system of Eqs.~(\ref{Subeq::q1dot})-(\ref{Subeq::phi1dot}) can be written as two second-order equations in~$\phi_k$ (\textit{de facto}, KCLs at nodes~$\Circled{1}$ and $\Circled{2}$ in the diagram of Fig.~\ref{Fig::figure01_res-coupl-circ-diagr_mt_mariantoni}):
\begin{widetext}
	\begin{subequations}
		\begin{empheq}[left=\empheqlbrace]{align}
			&L_1C_1\accentset{\bullet\bullet}{\phi}_1
			 +\dfrac{L_1}{R_{12}}\accentset{\bullet}{\phi}_1
			 +\phi_1
			 \quad
			 -\dfrac{L_1}{R_{12}}\accentset{\bullet}{\phi}_2
			 =0
			\label{Subeq::KCLp1p}
			\\[2mm]
			&-\dfrac{L_2}{R_{12}}\accentset{\bullet}{\phi}_1
			\quad
			+L_2C_2\accentset{\bullet\bullet}{\phi}_2
			+\dfrac{L_2}{R_{12}}\accentset{\bullet}{\phi}_2
			+\phi_2
			=0.
			\label{Subeq::KCLp2p}
		\end{empheq}
	\end{subequations}
\end{widetext}

The complex eigenfrequencies~$s$ are readily found with the usual ans\"{a}tze~$\phi_1\rightarrow\bar{\phi}_1e^{st}$ and $\phi_2\rightarrow\bar{\phi}_2e^{st}$ ($\bar{\phi}_1$ and $\bar{\phi}_2$ are complex amplitudes). Under these assumptions, the system of Eqs.~(\ref{Subeq::KCLp1p}) and (\ref{Subeq::KCLp2p}) becomes a system of two second-order algebraic equations. The system's determinantal equation allows us to obtain~$s_{\textrm{par}}^{\mp}=\mp i\omega_{\textrm{r}}$ and $s_{\textrm{symm}}^{\mp}=-\alpha\mp i\omega_{\textrm{r}}$, where the damping coefficient~$\alpha=1/\tau_{RC}$ and the time constant~$\tau_{RC}=R_{12}C_1$ (assuming for simplicity that~$C_1=C_2$). The symmetric eigenfrequencies~$s_{\textrm{symm}}^{\mp}$ resemble those of two lossless coupled pendulums; however, in our case the coupling contribution to the frequencies is due to the lossy part of the spring (i.e., the coupling resistance~$R_{12}$).

In the steady-state regime, we finally find the solutions as the linear superposition of the two normal eigenmodes:
\begin{widetext}
	\begin{subequations}
		\begin{empheq}{align}
			&\phi^{\textrm{st}}_1
			=
			\lim_{t\to+\infty}
			\left[
			\dfrac{A}{2}
			\cos(\omega_{\textrm{r}}t)+
			\dfrac{B}{2}
			e^{-\alpha t}\cos(\omega_{\textrm{r}}t)
			\right]
			=
			\dfrac{A}{2}
			\cos(\omega_{\textrm{r}}t)
			\label{Subeq::phi1st}
			\\[2mm]
			&\phi^{\textrm{st}}_2
			=
			\lim_{t\to+\infty}
			\left[
			\dfrac{A}{2}
			\cos(\omega_{\textrm{r}}t)-
			\dfrac{B}{2}
			e^{-\alpha t}\cos(\omega_{\textrm{r}}t)
			\right]
			=
			\dfrac{A}{2}
			\cos(\omega_{\textrm{r}}t),
			\label{Subeq::phi2st}
		\end{empheq}
	\end{subequations}
\end{widetext}
where the constants~$A$ and $B$ depend on the initial conditions. As expected, the steady-state solutions of a resistively coupled pair of resonators are those associated with the parallel normal modes.

\section{CONCLUSIONS}
	\label{Sec::CONCLUSIONS}

In our study, we develop a theory of linear and weakly nonlinear nonconservative electrical circuits grounded in the principles of Hamiltonian and Rayleigh dissipation function. This approach suggests that the resulting equations of motion can be described by Kirchhoff's laws, a remarkable finding that holds true in both classical and quantized scenarios.

In the classical realm, we show that the Poisson-Rayleigh brackets effectively carry out a derivative of the Hamiltonian or Rayleigh dissipation function with respect to the canonical coordinates. Similarly, in the quantized domain, we observe the same outcome using commutators.

By investigating electrical circuits, we leverage the dual aspect of Kirchhoff's voltage and current laws to both impose topological constraints and formulate the circuit's equations of motion. This methodology is applied to study quantum synchronization in various circuits, including pathological circuits with Hamiltonian singularities.

As our work is exclusively conducted in the Heisenberg picture, a potential expansion of our research would involve developing techniques to quantify quantum-mechanical properties such as entanglement. Our future endeavors will focus on devising methodologies that enable us to determine the density matrix starting from the system's observables in the Heisenberg picture--crucially--when nonconservative elements are present.

\appendix
\section{Poisson-Rayleigh Brackets}
	\label{App::Poisson-Rayleigh:Brackets}

Following the work of Ref.~\cite{Mariantoni:2021}, the classical Poisson~\cite{Cercignani:1976} brackets can be generalized to account for the first derivatives of the canonical coordinates. This method allows us to succinctly write the~EOMs of a given circuit, in presence of nonconservative elements. These elements are accounted for by means of the Rayleigh dissipation function, which depends on the coordinates' derivatives (i.e., voltages~$\accentset{\bullet}{\phi}_k$ and currents~$\accentset{\bullet}{q}_k$).

The~EOMs are found from
\begin{subequations}
	\begin{empheq}[left=\empheqlbrace]{align}
		\accentset{\bullet}{q}_k
		& = \{ ( \mathcal{H} , \mathcal{D} ) , ( q_k , \accentset{\bullet}{q}_k ) \} \nonumber\\
		& = \sum_{k=1,2} \left( \dfrac{\partial \mathcal{H}}{\partial q_k} \dfrac{\partial q_k}{\partial \phi_k} - \dfrac{\partial \mathcal{H}}{\partial \phi_k} \dfrac{\partial q_k}{\partial q_k} \right) \nonumber\\
		& + \sum_{k=1,2} \left( \dfrac{\partial \mathcal{D}}{\partial \accentset{\bullet}{q}_k} \dfrac{\partial \accentset{\bullet}{q}_k}{\partial \accentset{\bullet}{\phi}_k} - \dfrac{\partial \mathcal{D}}{\partial \accentset{\bullet}{\phi}_k} \dfrac{\partial \accentset{\bullet}{q}_k}{\partial \accentset{\bullet}{q}_k} \right)
		\label{Eq::Poisson:Rayleigh:qkdot}
		\\[2mm]
		\accentset{\bullet}{\phi}_k
		& = \{ ( \mathcal{H} , \mathcal{D} ) , ( \phi_k , \accentset{\bullet}{\phi}_k ) \} \nonumber\\
		& = \sum_{k=1,2} \left( \dfrac{\partial \mathcal{H}}{\partial q_k} \dfrac{\partial \phi_k}{\partial \phi_k} - \dfrac{\partial \mathcal{H}}{\partial \phi_k} \dfrac{\partial \phi_k}{\partial q_k} \right) \nonumber\\
		& + \sum_{k=1,2} \left( \dfrac{\partial \mathcal{D}}{\partial \accentset{\bullet}{q}_k} \dfrac{\partial \accentset{\bullet}{\phi}_k}{\partial \accentset{\bullet}{\phi}_k} - \dfrac{\partial \mathcal{D}}{\partial \accentset{\bullet}{\phi}_k} \dfrac{\partial \accentset{\bullet}{\phi}_k}{\partial \accentset{\bullet}{q}_k} \right)
		.
		\label{Eq::Poisson:Rayleigh:phikdot}
	\end{empheq}
\end{subequations}

\begin{acknowledgments}
This research was undertaken thanks in part to funding from the Canada First Research Excellence Fund~(CFREF). We acknowledge the support of the Natural Sciences and Engineering Research Council of Canada~(NSERC), [Application Number: RGPIN-2019-04022].
\end{acknowledgments}

\bibliography{Bibliography}

\end{document}